\documentclass[12pt]{article}

\usepackage[normalem]{ulem}
\usepackage[mathscr]{eucal}
\usepackage[version=3]{mhchem}

\usepackage{epsfig,latexsym,amsfonts,amsmath,amsthm,amssymb,amsbsy,multirow,slashed,wasysym,textcomp,dsfont,comment,mathtools,cancel,cite,datetime,appendix,BOONDOX-calo}
\usepackage{tocloft}
\usepackage{bbold}
\usepackage{tikz}
\usetikzlibrary{backgrounds}
\usetikzlibrary{patterns}
\usetikzlibrary{arrows.meta}
\usetikzlibrary{shapes,snakes}
\tikzstyle{bag} = [align=center]
\usetikzlibrary{decorations.pathmorphing}
\def\bea{\begin{eqnarray}}
\def\eea{\end{eqnarray}}

\usepackage{hyperref}
\usepackage{subcaption}

 \newcommand{\badat}{\begin{alignedat}}
 \newcommand{\eadat}{\end{alignedat}}
 
 \def\be{\begin{equation}}
\def\ee{\end{equation}}

\def\p{\partial}

\usepackage{xcolor}

\newcommand{\pink}[1]{\textcolor{\pink}{#1}}

\definecolor{dblue}{rgb}{0.2,0.50,0.80}

\def\O{\mathcal{O}}

\def\D{\mathcal{D}}

\def\bz{{\bar z}}
\def\bw{{\bar w}}

\def\bz{{\bar z}}
\def\bw{{\bar w}}

\DeclareFontFamily{OT1}{pzc}{}
\DeclareFontShape{OT1}{pzc}{m}{it}{<-> s * [1.10] pzcmi7t}{}
\DeclareMathAlphabet{\mathpzc}{OT1}{pzc}{m}{it}

\definecolor{vert}{rgb}{0.1367 0.543 0.1367}

\numberwithin{equation}{section} 

\begin{document}

 \begin{titlepage}
  \thispagestyle{empty}
  \begin{flushright}
  \end{flushright}
  \bigskip
  \begin{center}

        \baselineskip=13pt {\LARGE \scshape{
       A Comment on Loop Corrections \\[.75em]
       to the Celestial Stress Tensor}}
     
      \vskip1cm 

   \centerline{ 
   {Sabrina Pasterski}
}

\bigskip\bigskip

  \centerline{Perimeter Institute for Theoretical Physics, Waterloo, ON N2L 2Y5, Canada }

 \bigskip
 
 \centerline{Princeton Center for Theoretical Science, Princeton, NJ 08544, USA}

\bigskip\bigskip

\end{center}

\begin{abstract}
 \noindent 
In this note we show how the 1-loop exact correction to the subleading soft graviton theorem arising from IR divergences of scattering amplitudes matches onto the quadratic corrections to the soft charges computed from the BMS flux algebra. In the process, we examine how the BMS flux construction extends the celestial diamond framework to non-linear order and non-trivial vacua, and provides the natural symmetry generators for Celestial CFT.
  
\end{abstract}

\end{titlepage}

\tableofcontents

\section{Introduction}

Much of the recent progress towards constructing a flat space hologram~\cite{Pasterski:2021raf} has been guided by our understanding of the symmetries of the bulk and the behavior of bulk excitations near the conformal boundary~\cite{Strominger:2017zoo}. Efforts to pin down the relevant symmetry group have relied on input from both amplitudes and relativity communities. While it is perhaps not surprising in hindsight that the IR behavior of scattering~\cite{Weinberg:1965nx,Chung:1965zza,Kulish:1970ut} is imprinted on the large distance, long timescale modes on the boundary~\cite{Bondi:1962px,Sachs:1962wk,Sachs:1962zza}, 
the concrete proposal of an equivalence between asymptotic symmetries and soft theorems~\cite{Strominger:2013lka,Strominger:2013jfa} has provided an interesting avenue to explore extensions of the asymptotic symmetry group. In this story, the subleading soft graviton plays a particularly important role because it encodes a Virasoro symmetry of the $\mathcal{S}$-matrix~\cite{Barnich:2011ct,Barnich:2011mi,Cachazo:2014fwa,Kapec:2014opa}, motivating the celestial hologram~\cite{Pasterski:2019ceq,Raclariu:2021zjz,Pasterski:2021rjz, McLoughlin:2022ljp}.

The celestial holographic map looks at $\mathcal{S}$-matrix elements in a boost basis where they transform as correlation functions of primaries under the action of the Lorentz group~\cite{Pasterski:2016qvg,Pasterski:2017kqt}. For massless external states this is simply implemented by a Mellin transform
\be\label{mellin}
\langle \mathcal{O}^\pm_{\Delta_1}(z_1,\bz_1)...\mathcal{O}^\pm_{\Delta_n}(z_n,\bz_n)\rangle=\prod_{i=1}^n \int_0^\infty d\omega_i \omega_i^{\Delta_i-1} \langle out|\mathcal{S}|in\rangle
\ee
trading the energy for a boost weight. Coupling to gravity promotes these quasi-primaries to Virasoro primaries precisely because of the subleading soft graviton theorem. Taking this presentation of $\mathcal{S}$-matrix elements as a 2D system to heart and interpreting the (complexified) collinear limits of scattering as encoding the data of a radially quantized 2D celestial CFT unveils even richer symmetry structures~\cite{Guevara:2021abz,Strominger:2021lvk} than anticipated. Meanwhile, input from the twistor~\cite{Adamo:2021lrv,Costello:2022wso,Costello:2022upu} and relativity communities~\cite{Freidel:2021ytz} have provided critical insight into the expected scope.

This note focuses on the loop corrections to the subleading soft graviton. In particular, we show that insights from the leading conformally soft sector~\cite{Himwich:2020rro} dictate a slight modification to the 1-loop exact correction to the celestial stress tensor~\cite{He:2015zea} in a manner that matches onto the quadratic corrections to the soft charges computed from the BMS flux algebra in~\cite{Donnay:2021wrk}. Our motivation for doing so is two-fold.  On the one hand, it serves as a nice example of gainfully merging insights from the 2D organization of the conformally soft sector, IR divergences in amplitudes, and covariant phase space methods in relativity.   On the other, it provides an ideal opportunity to connect the BMS flux algebra statements to the celestial diamond framework~\cite{Pasterski:2021fjn,Pasterski:2021dqe} and its null state predecessors~\cite{Banerjee:2018fgd,Banerjee:2019aoy,Banerjee:2019tam}.

Let us take a moment to expand on this latter point in anticipation of a more detailed discussion below. The modus operandi is essentially the same principle motivating the celestial framework in the first place: organize the objects we care about into representations of the relevant symmetries. A surprising amount of mileage can be attained just using the Lorentz group.  Indeed, if we are looking at representations for the matter fields we want to couple to gravity, we will be interested in this group anyway.

In the celestial basis, our single particle states are in boost eigenstates. While our hologram is presented as two dimensional, the space of on-shell momenta is three dimensional. This manifests itself as continuous spectrum on the principal series~\cite{Pasterski:2017kqt}. To capture conformally soft theorems we need to analytically continue to general complex dimensions~\cite{Guevara:2019ypd,Puhm:2019zbl,Pate:2019mfs}.  This tunability allows us to go to conformal dimensions where our state has a primary descendant.  For example, a negative helicity spin-$s$ particle has primary descendants
\be
\mathcal{O}_{soft}=\frac{1}{k!}\p_\bw^k\mathcal{O}_{\Delta,-s},
\ee
at
\be
\Delta=1-s-k,
\ee
with similar expressions for the positive helicity sector. These are observed to decouple in scattering amplitudes for the known soft theorems.  For recent discussions on the tower of symmetry currents phrased in the celestial diamond language see \cite{Donnay:2022sdg,Freidel:2021ytz}. 

We will be interested solely in the leading and subleading soft graviton here. Now the Ward Identities for asymptotic symmetries~\cite{Strominger:2017zoo} take the following general form in $\mathcal{S}$ matrix insertions
\be\label{ward}
\langle out |Q^+[\xi]\mathcal{S}-\mathcal{S} Q^-[\xi]|in\rangle=0
\ee
where the charge contributions from $\mathcal{I}^\pm$ further split into a hard and a 
soft term
\be\label{splush}
Q^\pm[\xi]=Q_S^\pm[\xi]+Q_H^\pm[\xi].
\ee
The insight of~\cite{Banerjee:2018fgd,Banerjee:2019aoy,Banerjee:2019tam} is that for the soft and hard operators to add to zero they should have the same weights under the Lorentz group.  In practice, the constraint equations used to recast the standard canonical charges in to the form~\eqref{splush}  relate descendants of modes with different spins. Indeed, the candidate currents in CCFT arise from particular choices of the gauge parameter such that the soft operator has definite weights.  By the aforementioned argument, the hard particles coupling to this operator must also have these same weights. Indeed we see that while the radiative modes have spin-2, the sources producing the memory effect are captured by operators of lower spin. More generally, for a given matter operator one can ask what radiative modes it can couple to.  Similarly, for a given conformally soft mode with a primary descendant one can ask what operators can source it.  This latter route is relevant for assessing corrections to the soft theorems.

Now the hard operator is a primary but not a descendant. From this perspective, the global symmetries arise if there are gauge parameters such that the soft charge vanishes
\be
\sum_i \langle \mathcal{O}_1...\delta_i\mathcal{O}_i...\mathcal{O}_n\rangle=0.
\ee
This happens for gauge parameters in the kernel of the descendancy relation whenever the behavior of the soft operator in correlators is such that we can integrate by parts. In practice, one finds that this terminates at the usual Poincar\'e group for gravity or the standard global charge for gauge theory. Indeed, the majority of the asymptotic symmetries are spontaneously broken by a choice of vacuum. In this case, the soft charge operators shift the vacuum, as manifested in $\mathcal{S}$-matrix elements by the insertion of an extra incoming or outgoing soft gauge boson.

The upshot of the BMS flux algebra~\cite{Donnay:2021wrk} is to generalize this statement to construct objects that are covariant under the extended BMS group.  Namely, not only the total fluxes but also the hard and soft splittings are required to provide an appropriate representation of this larger group. In doing so they effectively extend the celestial diamond framework to non-linear order and non-trivial vacua.  These $u$-integrated operators provide the natural symmetry generators for Celestial CFT.   The slices of constant (spacelike) Rindler time intersect the conformal boundary on surfaces that run along the generators of $\mathcal{I}^\pm$. If we want to describe a state in the radial quantized CCFT defined on a contour on the celestial sphere, this lifts to a two dimensional slice of the boundary of this form, extending along the generators so as to precisely support the BMS fluxes~\cite{Pasterski:2022jzc}.   That we are able to get this mileage -- namely, celestial operators with the appropriate symmetry transformations that hold beyond leading order in perturbation theory -- from our understanding of the bulk equations of motion near the boundary is consistent with  either the celestial extrapolate dictionary~\cite{Pasterski:2021dqe,Donnay:2022sdg} or the Carrollian description~\cite{Donnay:2022aba,Bagchi:2022emh}. The fact that the flux algebras avoid certain issues with the charges associated to fixed  $u$-cuts serves as a nicer motivation for the CCFT version of the hologram.

This note is organized as follows.  In section~\ref{sec:flux}, we set up our conventions for the Bondi gauge metric near $\mathcal{I}^+$ and review the construction of the BMS fluxes from~\cite{Donnay:2021wrk}.  We then compare this to the loop correction to the celestial stress tensor proposed in~\cite{He:2017fsb} in section~\ref{sec:loop}. Here insights from the supertranslation vertex operators in~\cite{Himwich:2020rro} let us match to the the quadratic corrections from~\cite{Donnay:2021wrk}. We close with a recap of the assumptions and discussion of the takeaways in section~\ref{sec:discuss}.

\section{BMS Flux Algebra}\label{sec:flux}
In this section we review the construction of~\cite{Donnay:2021wrk}. The coadjoint representation of the BMS group was examined in~\cite{Barnich:2021dta}, and used to study the non-radiative phase space of asymptotically flat spacetimes. Various issues that arise when radiation is included can be avoided by looking at certain integrated quantities called BMS fluxes -- defined in terms of phase space data integrated along generators of null infinity rather than living at a fixed-$u$ cut~\cite{Campiglia:2020qvc,Compere:2020lrt}. 
Namely, one can define a flux algebra that closes under the standard bracket and is finite once we impose appropriate late-$u$ falloffs on our phase space. 

\subsection{A Bondi Gauge Primer}
Most of our notation for the Bondi gauge metric of asymptotically flat spacetimes will be familiar from~\cite{Strominger:2017zoo}. Near future null infinity, the metric in Bondi gauge takes the form~\cite{Bondi:1962px,Sachs:1962wk,Sachs:1962zza}
\be\badat{3}\label{bondi}
ds^2=&-du^2-2dudr+2r^2\gamma_{z\bz}dzd\bz+\frac{2m_B}{r}du^2+rC_{zz}dz^2+rC_{\bz\bz}d\bz^2\\
&+\bigl[(D^zC_{zz}-\frac{1}{4r}D_z(C_{zz}C^{zz})+\frac{4}{3r}N_z)dudz+c.c.\bigr]+...
\eadat\ee
For ease of comparing to the notation in~\cite{Donnay:2021wrk}, note that $\gamma_{z\bz}=(\Omega_S\bar\Omega_S)^{-1}$ is the unit round metric in projective coordinates.  We will see later that such factors are often used to construct objects with definite weights.  The latter notation avoids confusion over the index structure so we will adopt it here. The $u$-evolution of the Bondi mass and angular momentum aspect are controlled by the constraint equations $G_{u\mu}=8\pi G T^M_{u\mu}$ at large-$r$
\be\label{constraints}
\p_u m_B=\frac{1}{4}[D_z^2N^{zz}+D_\bz^2 N^{\bz\bz}]-T_{uu},~~~\p_u N_z=\frac{1}{4}D_z[D_z^2C^{zz}-D_\bz^2 C^{\bz\bz}]+D_zm_B-T_{uz}
\ee
 while the radiative data is captured by the news tensor $N_{zz}=\p_u C_{zz}$. In~\eqref{constraints} we've suggestively grouped terms quadratic in the metric into the $T_{\mu\nu}$ with the matter fields
\be\badat{3}
T_{uu}&\equiv \frac{1}{4}N_{zz}N^{zz}+4\pi G\lim\limits_{r\rightarrow\infty}r^2T^M_{uu}\\
T_{uz}&\equiv -\frac{1}{4}D_z[C_{zz}N^{zz}]-\frac{1}{2}C_{zz}D_z N^{zz}+8\pi G\lim\limits_{r\rightarrow\infty}r^2T^M_{uz},
\eadat\ee
incorporating the gravitational contribution to the the null energy and angular momentum, in what is treated as a `hard' term coupling to finite energy gravitons. In the remainder of this section we will stick with a purely gravitational theory.  As we will soon see, one of the nice upshots of~\cite{Donnay:2021wrk} is that they revisit the soft and hard splittings using the charge algebra as a guide.  This is very much in the spirit of the celestial diamonds~\cite{Pasterski:2021fjn,Pasterski:2021dqe} and null state predecessors~\cite{Banerjee:2018fgd,Banerjee:2019aoy,Banerjee:2019tam} outlined in the introduction.

Now this class of falloffs is preserved by the following asymptotic symmetries 
\be\badat{3}\label{xiy}
\xi & = f\p_u-\frac{1}{r}(D^zf \p_z+D^\bz f \p_\bz)+D^zD_z f\p_r \\
&~~~+(1+\frac{u}{2r})Y\p_z-\frac{u}{2r}D^\bz D_z Y\p_\bz-\frac{1}{2}(u+r)D_zY\p_r +\frac{u}{2}D_zY\p_u+c.c.+...
\eadat\ee
where $f(z,\bz)$ and $Y(z)$ depend only on the celestial sphere coordinates, and parameterize supertranslations and superrotations, respectively.  In what follows it will be convenient to strip out a factor of the round sphere metric from our supertranslation parameter
\be
f=(\Omega_S\bar\Omega_S)^{-\frac{1}{2}}T.
\ee 
The ellipses in~\eqref{xiy} indicates (field-dependent) subleading-in-$r$ terms that ensure these killing vectors preserve the Bondi gauge conditions.
Using the modified Lie bracket $[\xi_1,\xi_2]_*=[\xi_1,\xi_2]-\delta_{\xi_1}\xi_2+\delta_{\xi_2}\xi_1$~\cite{Barnich:2010eb} to take into account this field dependence, 
we get the expected closure of the commutation relations
\be
[\xi(T_1,Y_1,\bar{Y}_1),\xi(T_2,Y_2,\bar{Y}_2)]_*=\xi(T_{12},Y_{12},\bar{Y}_{12})
\ee
where
\be
T_{12}=Y_1\p T_2-\frac{1}{2}\p Y_1 T_2 -(1\leftrightarrow 2)+c.c.,~~~~Y_{12}=Y_1\p Y_2-(1\leftrightarrow 2),~~~\bar Y_{12}=\bar Y_1\p \bar Y_2-(1\leftrightarrow 2).
\ee
Here $\p$ is a partial derivative, in contrast to the $D_z$ appearing above, which is a covariant derivative with respect to the round sphere metric. 

Finite superrotations introduce defects into the round sphere metric~\cite{Compere:2016jwb,Adjei:2019tuj}.  Indeed, if we act on the usual Minkowski vacuum with such finite transformations we get the following `vacuum' shear tensor and News~\cite{Compere:2016jwb}
\be\label{lateC}
C^{vac}_{zz}=(u+C)\Theta_{zz}-2D_z^2C~~~ N^{vac}_{zz}=\p_u C^{vac}_{zz}=\Theta_{zz}
\ee
where $C$ is the Goldstone mode and $\Theta_{zz}$ is the superrotation Goldstone mode.  Namely, under infinitesimal transformations of the form~\eqref{xiy} these shift as follows
\bea
\label{thetatrafo}
\delta_f C =f,~~~\delta_Y\Theta_{zz} =\, Y \p_z \Theta_{zz}+ 2 \p_z Y \Theta_{zz} - \p^3_zY.
\eea
This makes it natural to define the `physical' news tensor~\cite{Compere:2018ylh}
\be\label{Nsub}
\hat{N}_{AB}=N_{AB}-N^{vac}_{AB}
\ee
which captures the (non-vacuum) radiative data and transforms homogeneously under the asymptotic symmetry group.

The early and late $u$ limits of radiative spacetimes reverts to vacua of this form where the mode $C$ can take different asymptotic values $C_{\pm}:=C(u\leftrightarrow\pm \infty)$. Meanwhile, the superrotation Goldstone mode  can be recast in terms of a Liouville stress tensor~\cite{Compere:2018ylh}
\be
\Theta_{zz}=\left[\frac{1}{2}D_A\Phi D_B\Phi-D_AD_B\Phi\right]^{TF}~~~
\Phi(z,\bz)=\varphi(z)+\bar{\varphi}(\bz)+\ln(\Omega_S\bar\Omega_S). 
\ee
For dynamical solutions $\Delta C=C_+-C_-$ encodes the memory effect and is symplectically paired with the Goldstone mode which we can identify with $C_-$. Similarly, the spin memory effect~\cite{Pasterski:2015tva} is measured by the celestial stress tensor and is symplectically paired with the superrotation Goldstone mode~\cite{Himwich:2019qmj}.  The naive over-counting of currents  we get when treating the Goldstone and memory modes as two independent fields can be used to our advantage.  Indeed, by taking certain linear combinations of  these  two (2,0) modes~\cite{Ball:2019atb,Nguyen:2022zgs}, one can define a sector of CCFT with a nontrivial central charge \cite{Pasterski:2022lsl}.

Now this Liouville mode also plays an important role in the construction of conformal fields on the celestial sphere. Consider a conformal transformation of the form $z\mapsto z'(z)$ and $\bz\mapsto \bz'(\bz)$ along with an appropriate Weyl rescaling.\footnote{From the point of the 2D celestial sphere, the Weyl rescaling comes from the action of the superrotations on the $r$ coordinate so that the leading round sphere metric is unchanged away from certain punctures where $Y$ has poles.} A conformal field of weight $(h,\bar h)$ transforms as follows
\be
\phi'_{h,\bar{h}}(x')=\left(\frac{\p z}{\p z'}\right)^h\left(\frac{\p \bz}{\p \bz'}\right)^{\bar h}\phi_{h,\bar{h}}(x)
\ee
where $h=\frac{1}{2}(\Delta+J)$ and $\bar h=\frac{1}{2}(\Delta-J)$.  By contrast, for SL$(2,\mathbb{C})$ primaries we restrict to the M\"obius transformations $z'=\frac{az+b}{cz+d}$ and similarly for $\bz$. Taking into account the transformation properties of the Liouville field
\be
\delta_{T,Y}\Phi=Y^AD_A\Phi+D_AY^A
\ee
one finds that the following Weyl covariant derivatives~\cite{Barnich:2021dta,Donnay:2021wrk}
\be
\D \phi_{h,\bar{h}}=[D_z-h\p \Phi]\phi_{h,\bar{h}},~~~\bar{\D} \phi_{h,\bar{h}}=[D_\bz-\bar{h}\bar{\p} \Phi]\phi_{h,\bar{h}}
\ee
map us to conformal primaries with shifted weights $(h+1,\bar h)$ and $(h,\bar h+1)$, respectively. These conformal derivatives also appear in \cite{Freidel:2021qpz,Freidel:2021dfs}. We will see that these derivatives neatly incorporate corrections to the soft charges when we look at scattering around non-trivially superrotated vacua. 

\subsection{BMS Fluxes}
The upshot of~\cite{Barnich:2021dta} is that the supertranslation parameter $T$ and superrotation parameters $Y$ and $\bar Y$ have definite conformal weights. The same is true of the supermomentum and super angular momentum operators which appear in the canonical charges. 
The authors of~\cite{Donnay:2021wrk} show that when one goes beyond the non-radiative sector these have a natural expression as BMS fluxes.  First, the supermomentum flux takes the form
\be
P=\frac{1}{4\pi G}\int du \p_u {\mathpzc M},~~~{\mathpzc M}=(\Omega_S\bar\Omega_S)^{-\frac{3}{2}}[m_B+\frac{1}{8}(C^{zz}N^{vac}_{zz}+C^{\bz\bz}N^{vac}_{\bz\bz})]
\ee
where the supermomentum $\mathpzc M$ reduces to the familiar Bondi mass at early and late times in the trivial superrotation vacuum.  This combination is selected for the fact that it transforms as a conformal field with weights $(h,\bar h)=(\frac{3}{2},\frac{3}{2})$.  Similarly the super angular momentum flux takes the form
\be\badat{3}
J=\frac{1}{8\pi G}\int du \p_u {\mathpzc N},~~{\mathpzc N}=(\Omega_S\bar\Omega_S)^{-1}&\Big[N_\bz-u(\Omega_S\bar\Omega_S)^{\frac{3}{2}}D_\bz{\mathpzc M}+\frac{1}{4}C_{\bz\bz}D_\bz C^{\bz\bz}+\frac{3}{16}D_\bz(C_{zz}C^{zz})\\
&~~+\frac{u}{4}D^z[(D_z^2-\frac{1}{2}N^{vac}_{zz})C^z_{\bz}-(D_\bz^2-\frac{1}{2}N^{vac}_{\bz\bz})C^\bz_{z}]\Big]
\eadat\ee
where the super angular momentum $\mathpzc N$ involves an appropriate Bondi mass subtraction of the angular momentum aspect $N_\bz$, familiar from the charge expressions of~\cite{Barnich:2011mi} used for the early computations of the tree level Virasoro symmetry~\cite{Kapec:2014opa} and spin memory effect~\cite{Pasterski:2015tva}, but now with the corrections designed so that this object transforms as a conformal field with weights $(h,\bar h)=(1,2)$.  A similar expression involving the opposite helicity sector gives $\bar J$.

A key result of~\cite{Donnay:2021wrk} is that these fluxes can be further split into a soft and a hard part 
\be
P=P_{soft}+P_{hard},~~~J=J_{soft}+J_{hard}
\ee
which separately transform as conformal fields
\be\badat{3}\label{TP}
\delta_{T,Y,\bar Y}P_{soft/hard}&=[Y\p +\bar Y\bar \p +\frac{3}{2}\p Y +\frac{3}{2}\bar \p \bar Y] P_{soft/hard} \\
\delta_{T,Y,\bar Y}J_{soft/hard}&=[Y\p +\bar Y\bar \p +\p Y +2\bar \p \bar Y] J_{soft/hard}+[\frac{1}{2}T\bar\p +\frac{3}{2}\bar\p T]P_{soft/hard}.
\eadat\ee
To see this, let us define the following soft radiative modes
\be
\mathcal{N}^{(0)}=\frac{1}{16\pi G}
\int_{-\infty}^\infty du (\Omega_S\bar{\Omega}_S)^{\frac{1}{2}} \hat{N}_{zz},~~~\mathcal{N}^{(1)}=\frac{1}{16\pi G}
\int_{-\infty}^\infty du (\Omega_S\bar{\Omega}_S)u \hat{N}_{zz}
\ee
which we know to have weights $(\frac{3}{2},-\frac{1}{2})$ and $(1,-1)$, respectively, precisely because in the celestial amplitudes dictionary these correspond to the leading and subleading (conformally) soft theorems, as well as the supertranslation Goldstone mode
\be
\mathcal{C}=(\Omega_S\bar{\Omega}_S)^{\frac{1}{2}}C_-
\ee
with weight $(-\frac{1}{2},-\frac{1}{2})$. In terms of these modes we have
\be
P_{soft}={\cal D}^2 \bar{\cal N}^{(0)}+\bar {\cal D}^2 {\cal N}^{(0)},~~~
J_{soft}=-\bar{\mathcal{D}}^3\mathcal{N}^{(1)}-\bar{\mathcal{D}}^3\mathcal{C}\mathcal{N}^{(0)}-3\bar{\mathcal{D}}^2\mathcal{C}\bar{\mathcal{D}}\mathcal{N}^{(0)}
\ee
while  
\be
P_{hard}=-\frac{1}{16\pi G}\int du (\Omega_S\bar\Omega_S)^{-\frac{3}{2}} \hat{N}_{zz}\hat{N}^{zz},
\ee
and
\be
J_{hard}=\frac{1}{8\pi G}\int du (\Omega_S\bar\Omega_S)^{-1}\left[\frac{3}{4}\check{C}^-_{\bz\bz} D_\bz\hat{N}^{\bz\bz}+\frac{1}{4} \hat{N}^{\bz\bz}D_\bz  \check{C}^-_{\bz\bz} +\frac{u}{4}D_\bz (\hat{N}^{\bz\bz}\hat{N}_{\bz\bz})\right]
\ee
where we've defined
\be
\check{C}^\pm_{AB}=C_{AB}-C^{vac,\pm}_{AB}.
\ee
Unlike for the news tensor~\eqref{Nsub} we have to make a choice of which (early or late time) vacuum to use to define a subtracted $C_{AB}$ that transforms homogeneously~\eqref{lateC}. This is because while typical scattering processes undergo non trivial supertranslation vacuum transitions, we are not allowing for superrotation vacuum transitions~\cite{Strominger:2016wns}. Exchanging $C_-\mapsto C_+$ throughout would not affect the charge algebra.   Meanwhile, the electric boundary conditions at early/late $u$ imply that the two terms in $P_{soft}$ are equal.  This equality is an echo of the shadow redundancy at the level of the conformal primary wavefunctions and corresponding soft theorems at these conformal dimensions.

As discussed in more detail in section~\ref{sec:discuss}, we see the celestial diamond structure appearing in the linear-order contribution to the soft term.  Meanwhile, the hard fluxes vanish on vacuum solutions. For the authors of~\cite{Donnay:2021wrk} this splitting is based on consistency with the symmetry algebra. Namely, for an appropriate complexified integration contour, used to isolate modes in a double Laurent expansion in $z$ and $\bz$, the transformation laws~\eqref{TP} imply that the smeared fluxes
 \be
 F_{T,Y,\bar Y}=\int_S\frac{dzd\bz}{(2\pi i)^2}[TP+Y\bar J+\bar Y J]
 \ee
and their soft and hard splittings obey the following algebra~\cite{Barnich:2021dta,Compere:2020lrt,Campiglia:2021bap}
\be
\{F^{soft}_{(T_1,Y_1,\bar Y_1)},F^{hard}_{(T_2,Y_2,\bar Y_2)}\}=0,~~~\{F^{soft/hard}_{(T_1,Y_1,\bar Y_1)},F^{soft/hard}_{(T_2,Y_2,\bar Y_2)}\}=-F^{soft/hard}_{[(T_1,Y_1,\bar Y_1),(T_2,Y_2,\bar Y_2)]}
\ee
where the bracket $\{\cdot,\cdot\}$ is defined by 
\be
\{F_{(T_1,Y_1,\bar Y_1)},F_{(T_2,Y_2,\bar Y_2)}\}=\delta_{(T_1,Y_1,\bar Y_1)}F_{(T_2,Y_2,\bar Y_2)}.
\ee
These smeared operators naturally appear in the construction of the celestial symmetry generators. For what follows we will focus on the superrotation case.  

The authors of~\cite{Donnay:2021wrk} construct the following candidate celestial stress tensor from the soft super angular momentum flux as follows
\be
T(z)=\oint\frac{d\bz}{2\pi i}\bar{{J}}_{soft}(z,\bz),~~~\bar T(\bz)=\oint\frac{dz}{2\pi i}{{J}}_{soft}(z,\bz).
\ee
Examining the expression for $J_{soft}$ above, we see that the linearized piece matches the tree level stress tensor of~\cite{Kapec:2016jld}, while the quadratic terms almost (but not quite) match the loop level corrections found in~\cite{He:2017fsb}.  
Here, these terms were necessary for the superrotation generators to have the correct transformation properties under the action of supertranslations.
We will now see that we can readily resolve this tension by taking into account the soft dressings~\cite{Himwich:2020rro}.

\section{A 1-Loop Correction to the Stress Tensor}\label{sec:loop}

The subleading soft graviton theorem~\cite{Cachazo:2014fwa} gets a one loop exact correction which was recast as a modification to the celestial stress tensor~\cite{Kapec:2014opa,Kapec:2016jld} in~\cite{He:2017fsb}. Here we will see that this IR divergent part nicely matches onto the form of the quadratic corrections to the soft flux found in~\cite{Donnay:2021wrk}. Our discussion in this section will follow the notation of~\cite{He:2017fsb} and we will work in momentum space. Since these computations involve scattering around the trivial superrotation vacuum we will want to match onto the $N^{vac}_{AB}=0$ limit of the expressions in the previous section, for which $\hat{N}_{AB}\mapsto N_{AB}$ and ${\cal D}\mapsto D_z$. 

\subsection{Reviewing the Loop Corrected Soft Graviton}
Let us start by considering two $\cal S$-matrix elements:  ${\cal A}_n$ corresponding to the scattering of $n$ particles of arbitrary species and energies; and ${\cal A}^\pm_{n+1}$ describing the process with an additional soft graviton of helicity $\pm2 $ added to the out state
\be
{\cal A}_n=\langle out|{\cal S}|in\rangle,~~~{\cal A}^\pm_{n+1}=\langle out|a_\pm(k){\cal S}|in\rangle,
\ee
where $k=\omega q$ for some reference null vector ${q}$. At tree level the leading terms in an expansion as $\omega\rightarrow0$ take the following form 
\be\label{eq:softtheoremA}
\mathcal{A}^\pm_{n+1}(k^\mu=\omega q^\mu)= [S_n^{(0)\pm}+S_n^{(1)\pm}+\O(\omega)]{\cal A}_n
\ee
where
\be\label{softgrav}
S_{n}^{(0)\pm}=\frac{\kappa}{2}\sum_{k=1}^n\frac{p_{k\mu}p_{k\nu} \varepsilon^{\pm\mu\nu}}{\omega p_k\cdot q}\,,~~~S_n^{(1)\pm}=-i\frac{\kappa}{2}\sum_{k=1}^n\frac{p_{k\mu}\varepsilon^{\pm\mu\nu}q^\lambda J_{k\lambda\nu}}{p_k\cdot q}\,
\ee
and $\kappa=\sqrt{32\pi G}$. The saddle point approximation connect fields near null infinity to $\cal{S}$-matrix elements~\cite{He:2014laa}.  As discussed further in~\cite{Pasterski:2021dqe,Donnay:2022sdg}, this underpins the extrapolate dictionary of CCFT. For the case of the leading and subleading soft graviton of interest here, we have the following relation between (appropriately regulated) $u$-integrals of the news and the soft graviton modes near $\mathcal{I}^+$
\be
N^{(0)}_{\bz\bz}=\int du N_{\bz\bz}=-\frac{\kappa}{8\pi}\hat{\varepsilon}_{\bz\bz}\lim\limits_{\omega\rightarrow0}\omega [ a_-(\omega q)+a_+(\omega q)^\dagger]
\ee
and
\be
N^{(1)}_{\bz\bz}=\int du u N_{\bz\bz}=\frac{i\kappa}{8\pi}\hat{\varepsilon}_{\bz\bz}\lim\limits_{\omega\rightarrow0}(1+\omega\p_\omega)[ a_-(\omega q)- a_+(\omega q)^\dagger]
\ee
where these operators live at the point $\hat{q}$ on the celestial sphere, corresponding to the direction of the three momentum of the soft graviton.  We see that the $\omega$-dependent prefactors select precisely the leading and subleading soft theorems in these limits. The notation in \cite{Donnay:2021wrk} is chosen to match the ones here.  For the perturbative $\cal S$-matrix we often consider the trivial superrotation background (or infinitesimal superrotations around this $N^{vac}_{\bz\bz}=0$ vacuum), so it is indeed appropriate to promote $N_{\bz\bz}\rightarrow\hat{N}_{\bz\bz}$ in the expressions here when we want to consider scattering atop a non-trivial superrotation.

Let us now turn our focus to the subleading case. The tree level celestial stress tensor is defined as follows
\be \label{Ttree}
T_{zz}^{tree}
=\frac{4i}{\kappa^2}\int d^2 w\frac{\gamma^{w\bw}}{z-w}D_w^3N^{(1)}_{\bw\bw}.
\ee
While integrating by parts on the celestial sphere turns this into a shadow of the subleading soft graviton, in the language of the celestial diamonds we see that as written~\eqref{Ttree} is a lift of the  level-3 primary descendant.  This form matches the starting point of~\cite{Donnay:2021wrk}, modulo the analytic continuation in their integration contour.

While the leading soft graviton theorem is not corrected at loop order, the subleading soft graviton receives a one loop exact correction.  The authors of~\cite{He:2017fsb} focus on the IR divergent part which in dim-reg contributes  the following pole
\be
{\cal A}_n^{(1)}|_{div}=\frac{\sigma_n}{\epsilon}{\cal A}_n^{(0)}
\ee
where  $d=4-\epsilon$, ${\cal A}_n^{(\ell)}\kappa^{2\ell}$ is the contribution from $\ell$-th loop order, and
\be\label{sign}
\sigma_n=-\frac{1}{4(4\pi)^2}\sum_{i,j=1}^{n}(p_i\cdot p_j)\log\frac{\mu^2}{-2p_i\cdot p_j}.
\ee
With the hindsight of~\cite{Himwich:2020rro}, we can quickly recognize this Eikonal factor as coming from a correlation function of vertex operators constructed from the supertranslation Goldstone mode.  Since this perspective modifies the final form of the answer in~\cite{He:2017fsb}, it is worthwhile to first work out a few more of the steps in their original derivation.

Combining this loop correction factor with the soft limit of the $n+1$ particle amplitude~\eqref{eq:softtheoremA} gives
\be
\mathcal{A}^{(1)\pm}_{n+1}\big |_{div}\overset{\omega\rightarrow0}{\longrightarrow} \frac{\sigma_{n+1}}{\epsilon}[S_n^{(0)\pm}+S_n^{(1)\pm}]{\cal A}_n^{(0)}
\ee
where we are dropping terms of order $\mathcal{O}(\omega)$.  Now $\sigma_{n+1}$ contains terms that scale like $\mathcal{O}(\omega^0)$ when soft gravitons are exchanged between two hard legs and a part that scales like $\mathcal{O}(\omega)$ when one of the particles is the soft graviton
\be\label{signprime}
\sigma_{n+1}=\sigma_n+\sigma_{n+1}',~~~\sigma_{n+1}'=-\frac{1}{2(4\pi)^2}\sum_{i=1}^{n}\omega( p_i\cdot q)\log\frac{\mu^2}{-2p_i\cdot q}.
\ee 
The $\log \omega$ term coming from the $\ln p_i\cdot k$ vanishes by momentum conservation.  Regrouping terms we find that 
\be\label{eq:softdiv}
\mathcal{A}^{(1)\pm}_{n+1}\big |_{div}\overset{\omega\rightarrow0}{\longrightarrow} [S_n^{(0)\pm}+S_n^{(1)\pm}]{\cal A}_n^{(1)}\big|_{div}+\frac{1}{\epsilon}\Big(\sigma_{n+1}' S_n^{(0)\pm}-[S_n^{(1)\pm},\sigma_n]\Big){\cal A}_n^{(0)}
\ee
where the last term comes from commuting the dressing factor $\sigma_n$ through the subleading soft operator. The first term just accounts for the Eikonal exchanges among the hard particles, as expected at loop order, while the term in brackets scales like $\mathcal{O}(\omega^0)$ and corrects the subleading soft theorem.

Thus we've seen that the subleading soft graviton insertion gets a loop correction. The fact that the tree level soft theorem could be interpreted as a 2D stress tensor Ward identity came from how the original soft factor $S_n^{(1)}$ acts on the external states.  We can thus restore this Ward identity at loop level by moving the second set of terms in~\eqref{eq:softdiv} to the left hand side, shifting the tree level stress tensor via
\be\label{Tloop}
T_{zz}=T^{tree}_{zz}+T^{loop}_{zz}=\frac{4i}{\kappa^2}\int d^2 w\frac{\gamma^{w\bw}}{z-w}D_w^3\left(N^{(1)}_{\bw\bw} - N^{(1)}_{\bw\bw}\big|_{div} \right)
\ee
where
\be\label{n1div}
\langle out|N^{(1)}_{\bw\bw}{\cal S}|in\rangle\big|_{div} = \frac{i\kappa^3}{8\pi}\hat{\varepsilon}_{\bw\bw}\lim\limits_{\omega\rightarrow0}(1+\omega\p_\omega)\frac{1}{\epsilon}\Big(\sigma_{n+1}' S_n^{(0)-}-[S_n^{(1)-},\sigma_n]\Big)\langle out|{\cal S}|in\rangle.
\ee
Next, to connect to the BMS flux algebra, we want to rewrite this loop correction in terms of the metric data at null infinity. This turns out to be quadratic in the zero modes.  We will first follow the steps in the appendix of~\cite{He:2017fsb}, then revisit this computation using the insights of~\cite{Himwich:2020rro}, before finally comparing to~\cite{Donnay:2021wrk}.

To evaluate identify the appropriate operator that gives rise to this soft theorem, the authors started with the explicit momentum space realization of the correction term.  The $\sigma_n$ and $\sigma_{n+1}'$ are given in~\eqref{sign} and~\eqref{signprime} above, while the action of the soft factors can be similarly evaluated noting that the angular momentum operator in~\eqref{softgrav} is simply the following differential operator on a function of the external momenta
\be
J_{k\lambda\nu}=-i\left[p_{k\lambda}\frac{\p}{\p p_k^\nu} -p_{k\nu}\frac{\p}{\p p_k^\lambda} \right].
\ee
The combination of terms in~\eqref{n1div} explicitly evaluates to
\be\badat{3}\label{n1div}
&\frac{1}{\epsilon}\Big(\sigma_{n+1}' S_n^{(0)\pm}-[S_n^{(1)\pm},\sigma_n]\Big)\\
&=\frac{\kappa}{4(4\pi)^2\epsilon}\sum_{i,j=1}^n\left[\frac{(p_i\cdot \varepsilon^-)^2}{p_i\cdot q}(p_j\cdot q)\log\frac{p_j\cdot q}{p_i\cdot p_j}-(p_i\cdot \varepsilon^-)(p_j\cdot \varepsilon^-)\log\frac{\mu^2}{-2p_i\cdot p_j}\right]
\eadat\ee
in ${\cal S}$-matrix elements. After evaluating the appropriate descendants, the authors of~\cite{He:2017fsb} match their answer to the form of the leading soft factor to conclude that
\be\label{Tfirst}
T_{zz}=\frac{4i}{\kappa^2}\int d^2 w\frac{\gamma^{w\bw}}{z-w}\left[D_w^3N^{(1)}_{\bw\bw} -\frac{i}{2\pi\epsilon}\left(2N_{ww}^{(0)}D_wN_{\bw\bw}^{(0)}+D_w(N_{ww}^{(0)}N^{(0)}_{\bw\bw} )\right)\right]
\ee
reproduces this loop correction in ${\cal S}$-matrix insertions. The details can be found in Appendix A of that reference. While plugging in the soft theorem indeed matches the momentum space form of this loop correction, there is a slight ambiguity that explains the mismatch with~\cite{Donnay:2021wrk} where one of the memory modes is instead replaced by a supertranslation Goldstone mode. That replacement has a non-trivial effect on the flux algebra since we need a mode that shifts inhomogeneously in the supertranslations to reproduce the expected transformation~\eqref{TP}. 

\subsection{The Soft S-Matrix}
We will now show that if we take a step back and look at the IR dressing factors responsible for this divergent loop correction and how they are captured by vertex operators in the CCFT constructed from the celestial Goldstone modes~\cite{Himwich:2020rro}, we naturally resolve the tension between the results in~\cite{He:2017fsb} and~\cite{Donnay:2021wrk} and avoid an explicit appearance of the IR regulator in the answer.  Moreover, this type of ambiguity at the level of matching correlators is closely tied to the naive over-counting of currents seen in~\cite{Nande:2017dba,Pasterski:2022lsl}.  As in those studies, restricting to sectors where the Goldstone and memory modes are proportional to each other (in a manner fixed by consistency with the symmetries) reduces this over-counting and gives rise to non-trivial levels~\cite{Nande:2017dba} and central charges~\cite{Pasterski:2022lsl}. Our conventions in this subsection will follow~\cite{Himwich:2020rro}.  While we are still in momentum space here, a celestial basis version has been examined in~\cite{Arkani-Hamed:2020gyp}. For additional discussions of celestial models that capture the soft dynamics associated to these supertranslation Goldstone modes see~\cite{Kapec:2021eug, Nguyen:2021ydb}.

In four dimensions, virtual soft graviton exchanges between the undressed single particle states generate IR divergences~\cite{Weinberg:1965nx}.  This leading Eikonal behavior exponentiates in a way that can be factored out from the IR finite part of the amplitude 
\be\label{Sdressed}
\langle out |{\cal S}|in\rangle=\exp\left[\frac{1}{\epsilon} \frac{G}{\pi} \sum_{i,j=1}^n p_i\cdot p_j \ln \left(\frac{\mu^2}{-2p_i\cdot p_j}\right)\right]\widehat{\langle out |{\cal S}|in\rangle}.
\ee
Now this IR divergent piece has a natural interpretation in terms of the asymptotic symmetries. First, the vanishing of exclusive amplitudes when we remove the IR cutoff is a symptom of charge non-conservation for processes that do not include the appropriate soft radiation~\cite{Kapec:2017tkm}.  Moreover, the Minkowski vacuum spontaneously breaks the 
supertranslation symmetry and the dynamics of this IR sector is controlled by this symmetry breaking. 

The IR divergent piece in~\eqref{Sdressed} can be recast in terms of a correlator of Wilson lines. Keeping the celestial framework in mind, the authors of~\cite{Himwich:2020rro} construct gravitational Wilson line-like operators that take the form of vertex operators in the celestial CFT
\be
{ W}_k=e^{i\eta_k \omega_k C(z_k,\bz_k)}
\ee
where $p_k=\eta_k\omega_k q_k$ for a null reference vector $q_k$ parameterizing points on the celestial sphere and $C$ is the supertranslation Goldstone mode. We can `undress' the operators creating single particle external states
\be\label{Odress}
{\cal O}_k={W}_k \tilde  {\cal O}_k.
\ee
The inhomogeneous shift $\delta_f C=f$ translates to the following transformation of the vertex operators
\be
\delta_f {W}_k(z_k,\bz_k)=i\eta_k \omega_k f(z_k,\bz_k) { W}_k(z_k,\bz_k)
\ee
so that the dressed operators $\tilde{\cal O}_k$ are supertranslation invariant. Mapping the external states to correlators on the celestial sphere, we see that~\eqref{Sdressed} can be recast into the form
\be
\langle {\cal O}_1\cdots {\cal O}_n\rangle=\langle W_1\cdots W_n\rangle \langle \tilde{\cal O}_1\cdots\tilde{\cal O}_n\rangle.
\ee
Namely
\be\label{Wdressed}
\langle W_1\cdots W_n\rangle=\exp\left[-\frac{1}{2}\sum_{i\neq j}^n\eta_i\eta_j\omega_i\omega_j\langle C(z_i,\bz_i)C(z_j,\bz_j)\rangle\right]
\ee
where 
\be\label{Cdressed}
\langle C(z_i,\bz_i)C(z_j,\bz_j)\rangle =\frac{1}{\epsilon}\frac{G}{\pi}\frac{|z_{i}-z_j|^2}{(1+z_i\bz_i)(1+z_j\bz_j)}\ln  \frac{|z_i-z_j|^2}{(1+z_i\bz_i)(1+z_j\bz_j)}
\ee 
and we've used momentum conservation to match~\eqref{Sdressed} and~\eqref{Wdressed}-\eqref{Cdressed}. This takes the form familiar from the IR divergent loop correction~\eqref{sign} we encountered in the previous section.  Namely
\be
\frac{\sigma_n}{\epsilon}=\frac{1}{2\kappa^2}\eta_i\eta_j \omega_i\omega_j \langle C(z_i,\bz_i)C(z_j,\bz_j)\rangle,
\ee
again using total momentum conservation.

For the purpose at hand, the point we care about is that the correlators of two Goldstone modes versus the correlator of a Goldstone mode and the memory mode of the same weight in the symplectically paired celestial diamond have the kinematic structure~\cite{Pasterski:2021dqe}. Therefore, one needs to take care when matching expressions for the soft factors to deduce which boundary limits of bulk operators they correspond to.  This is phrased as an off diagonal level structure for the Kac-Moody like symmetry which on the flat celestial sphere takes the form
\be
P_zP_w~0,~~~P_z\tilde{P}_w\sim\frac{1}{(z-w)^2},~~~\tilde{P}_z\tilde{P}_w\sim -\frac{1}{\epsilon}\frac{G}{\pi}\frac{\bz-\bw}{z-w}
\ee
reminiscent of the gauge theory case~\cite{Nande:2017dba} (but where the signed-energies $\eta_k\omega_k$ play the role of the charge for the corresponding vertex operators $W_k$)
generated by the operators
\be
P_z=\frac{1}{4G}D^z N^{(0)}_{zz},~~~\tilde{P}_z=iD_z C
\ee
with weights $(h,\bar h)=(\frac{3}{2},\frac{1}{2})$ and $(\frac{1}{2},-\frac{1}{2})$ respectively.
The memory effect is measured by insertions of $P_z$ precisely because it reproduces the soft theorem when it contracts with one of the $C$ operators in the dressings $W_k$
\be
\langle P_z W_1\cdots W_n\rangle =\left[\sum_{j=1}^n i\eta_j\omega_j \langle P_z C(z_j,\bz_j)\rangle\right]\langle  W_1\cdots W_n\rangle 
\ee
where
\be
i\eta_j\omega_j \langle P_z C(z_j,\bz_j)\rangle=\frac{\eta_j\omega_j}{z-z_j}=-\frac{1}{\kappa}D^z \hat{\epsilon}_{zz} S^{(0)}_{n,j}
\ee
and we recognize the contribution from particle $j$ to the leading soft factor in~\eqref{softgrav}. However, upon taking derivatives we find
\be
D^z D_{z}^2 \langle C(z,\bz) C(z_j,\bz_j)\rangle =-\frac{i}{2\pi\epsilon}D^z  \langle N^{(0)}_{zz}(z,\bz) C(z_j,\bz_j)\rangle .
\ee
We see that this provides the relative factor in~\eqref{Tfirst} so that the following expression
\be\label{Tnew}
T_{zz}=\frac{4i}{\kappa^2}\int d^2 w\frac{\gamma^{w\bw}}{z-w}\left[D_w^3N^{(1)}_{\bw\bw} +\left(2D_w^2 C D_w N_{\bw\bw}^{(0)}+D_w(D_w^2 C N^{(0)}_{\bw\bw} )\right)\right]
\ee
similarly reproduces the momentum space form of the loop corrections.  This matches the expression from~\cite{Donnay:2021wrk} in the trivial superrotation vacuum. Moreover, it is consistent with the fact that we now see that $\sigma_n$ is coming from the $C$ correlators that capture Eikonal exchanges, and that these Eikonial terms multiply copies of the negative helicity soft theorems.  

\section{Discussion}\label{sec:discuss}
Let's now recap the highlights of what was needed for each of the computations we've combined and revisit the larger themes motivating our investigations. 

In the BMS flux computations~\cite{Donnay:2021wrk}, the authors stick to the purely gravitational theory but keep track of quadratic corrections that give rise to both the gravitational contribution to the hard charge and corrections to the soft operators.  If we want to add matter back in, the constraint equations would imply additional contributions to the hard fluxes.  The splitting into soft and hard is based on demanding that each contribution transforms appropriately under the extended BMS group.  The hard part is quadratic in the radiative modes, vanishes on the vacuum solutions, and is a conformal primary that is not itself a (Weyl covariant) primary descendant of other metric fields. The soft part reduces to a primary descendant in the linearized limit, matching the tree level proposal for a celestial stress tensor in~\cite{Kapec:2016jld}.  Meanwhile, certain quadratic corrections are necessary to reproduce the correct covariance properties.  Furthermore, this derivation extends previous construction to nontrivial superrotation vacua.

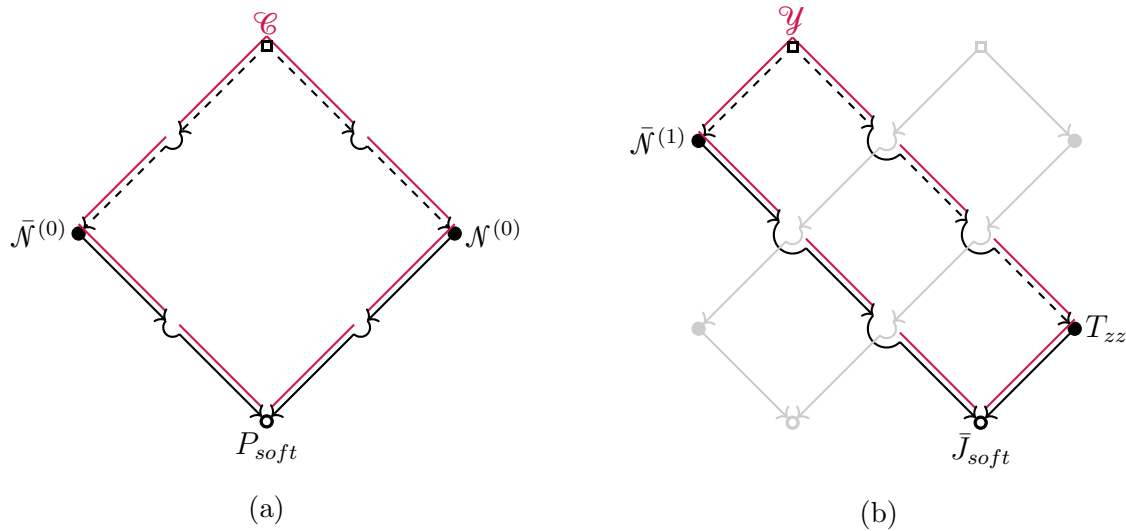
\begin{figure}[htb!]
\begin{subfigure}{.45\linewidth}
\centering
\begin{tikzpicture}[scale=1.25]
\definecolor{darkred}{rgb}{0.8, 0.1, 0.3}
\filldraw[black] (-2,1) circle (2pt)  node [left]{$\bar {\cal N}^{(0)}$};
\filldraw[black] (2,1) circle (2pt) node [right]{$ {\cal N}^{(0)}$};
\draw[thick] (1+.1414/2,0+.1414/2) arc (45:-135:.1);
\draw[thick] (-1+.1414/2,2+.1414/2) arc (45:-135:.1);
\draw[thick,dashed] (0,3) node [above,darkred]{${\cal C}$} ;
\draw[thick,darkred] (0,3+.1) --  (1-.1414/2,2+.1414/2+.1);
\draw[->,thick,dashed] (0,3) --  (1-.1414/2,2+.1414/2) node [above=4mm]{
};
\draw[thick,darkred] (1+.1414/2,2-.1414/2+.1) -- (2-.05,1+.05+.1);
\draw[->,thick,dashed] (1+.1414/2,2-.1414/2) -- (2-.05,1+.05) node [above=4mm]{
};
\draw[thick,darkred] (0,3+.1) --  (-1+.07,2+.07+.1);
\draw[->,thick,dashed] (0,3) --  (-1+.07,2+.07) node [above=4mm]{
};
\draw[thick,darkred] (-1-.07,2-.07+.1)-- (-2+.05,1+.05+.1);
\draw[->,thick,dashed] (-1-.07,2-.07)-- (-2+.05,1+.05) node [above=4mm]{
};
\draw[thick,darkred] (1-.07,0-.07+.1)-- (0+.05,-1+.05+.1) ;
\draw[->,thick] (1-.07,0-.07)-- (0+.05,-1+.05) ;
\node[fill=black,regular polygon, regular polygon sides=4,inner sep=1.6pt] at (0,3) {};
\node[fill=white,regular polygon, regular polygon sides=4,inner sep=.8pt] at (0,3) {};
\filldraw[black] (0,-1) circle (2pt)  node [below]{$P_{soft}$};
\filldraw[white] (0,-1) circle (1pt);
\draw[thick,darkred] (2,1+.1)  --  (1+.07,0+.07+.1) ;
\draw[->,thick] (2,1) node [below=4mm]{
} --  (1+.07,0+.07) node [below=4mm]{
} ;
\draw[thick,darkred] (-2,1+.1)  --  (-1-.1414/2,0+.1414/2+.1) ;
\draw[->,thick] (-2,1) node [below=4mm]{
} --  (-1-.1414/2,0+.1414/2) node [below=4mm]{
};
\draw[thick] (-1-.1414/2,0+.1414/2)  arc (135:315:.1);
\draw[thick,darkred] (-1+.1414/2,0-.1414/2+.1)  -- (0-.05,-1+.05+.1)  ;
\draw[->,thick] (-1+.1414/2,0-.1414/2)  -- (0-.05,-1+.05)  ;
\draw[thick] (1-.1414/2,2+.1414/2) arc (135:315:.1);
\end{tikzpicture}
\caption{}
\end{subfigure}
\begin{subfigure}{.45\linewidth}
\centering
\begin{tikzpicture}[scale=1.25]
\definecolor{darkgreen}{rgb}{.0, 0.5, .1};
\definecolor{darkgreen}{rgb}{.8, 0.8, .8};
\definecolor{darkred}{rgb}{0.8, 0.1, 0.3}
\definecolor{blue}{rgb}{0, 0, 0};
\filldraw[blue] (-2,2) circle (2pt) ;
\filldraw[blue] (2,0) circle (2pt)  node [right,black]{$T_{zz}$};
\filldraw[blue] (1,-1) circle (2pt) node [below,black]{$\bar J_{soft}$};
\filldraw[white] (1,-1) circle (1pt) ;
\filldraw[darkgreen] (2,2) circle (2pt)  node [right,black]{};
\filldraw[darkgreen] (-2,0) circle (2pt) node [left,black]{};
\filldraw[darkgreen] (-1,-1) circle (2pt) node [below,black]{};
\filldraw[white] (-1,-1) circle (1pt) ;
\draw[thick,darkred] (2,0+.1)-- (1+.05,-1+.05+.1);
\draw[thick,->,blue] (2,0)-- (1+.05,-1+.05);
\draw[thick,darkred] (-1,3+.1)-- (-2+.05,2+.05+.1);
\draw[thick,->,blue,dashed] (-1,3)-- (-2+.05,2+.05);
\draw[thick,darkgreen] (0+.1414/2,0+.1414/2) arc (45:-135:.1);
\draw[thick,darkgreen] (1+.1414/2,1+.1414/2) arc (45:-135:.1);
\draw[thick,darkgreen] (-1+.1414/2,1+.1414/2) arc (45:-135:.1);
\draw[thick,darkgreen] (0+.1414/2,2+.1414/2) arc (45:-135:.1);
\draw[thick,blue] (0-.1414,0+.1414) arc (135:315:.2);
\draw[thick,blue] (1-.1414,1+.1414) arc (135:315:.2);
\draw[thick,blue] (0-.1414,2+.1414) arc (135:315:.2);
\draw[thick,blue] (-1-.1414,1+.1414) arc (135:315:.2);
\draw[thick,darkred] (-2,2+.1)--  (-1-.1414,1+.1414+.1);
\draw[->,thick,blue] (-2,2)  node [left,black]{$\bar {\cal N}^{(1)}$} --  (-1-.1414,1+.1414);
\draw[thick,darkred] (-1,3+.1)  --  (0-.1414,2+.1414+.1);
\draw[->,thick,blue, dashed] (-1,3) node [above,darkred]{${\cal Y}$}  --  (0-.1414,2+.1414);
\draw[thick,darkred] (-1+.1414,1-.1414+.1) --  (0-.1414,0+.1414+.1);
\draw[->,thick,blue] (-1+.1414,1-.1414) --  (0-.1414,0+.1414);
\draw[thick,darkred] (0+.1414,0-.1414+.1) -- (1-.05,-1+.05+.1);
\draw[->,thick,blue] (0+.1414,0-.1414) -- (1-.05,-1+.05);
\draw[thick,darkred] (0+.1414,2-.1414+.1) --  (1-.1414,1+.1414+.1);
\draw[->,thick,blue,dashed] (0+.1414,2-.1414) --  (1-.1414,1+.1414);
\draw[thick,darkred] (1+.1414,1-.1414+.1) -- (2-.05,0+.05+.1);
\draw[->,thick,blue,dashed] (1+.1414,1-.1414) -- (2-.05,0+.05);
\draw[->,thick,darkgreen] (1,3) node [above,black]{} -- (2-.05,2+.05);
\draw[->,thick,darkgreen] (1,3) -- (0+.07,2+.07);
\draw[->,thick,darkgreen] (-2,0) -- (-1-.05,-1+.05);
\draw[->,thick,darkgreen] (2,2) -- (1+.07,1+.07);
\draw[->,thick,darkgreen] (1-.07,1-.07) -- (0+.07,0+.07);
\draw[->,thick,darkgreen] (0-.07,2-.07) -- (-1+.07,1+.07);
\draw[->,thick,darkgreen] (-1-.07,1-.07) -- (-2+.05,0+.05);
\draw[->,thick,darkgreen] (0-.07,0-.07) -- (-1+.05,-1+.05);
\node[fill=blue,regular polygon, regular polygon sides=4,inner sep=1.6pt] at (-1,3) {};
\node[fill=white,regular polygon, regular polygon sides=4,inner sep=.8pt] at (-1,3) {};
\node[fill=darkgreen,regular polygon, regular polygon sides=4,inner sep=1.6pt] at (1,3) {};
\node[fill=white,regular polygon, regular polygon sides=4,inner sep=.8pt] at (1,3) {};
\end{tikzpicture}
\caption{}
\end{subfigure}
\caption{Goldstone (red) and memory (black) diamonds for the leading (a) and subleading (b) soft graviton theorem.}
\label{fig:gravitondiamond}
\end{figure}

To compare these expressions with the amplitudes computations~\cite{Kapec:2014opa,Kapec:2016jld,He:2017fsb}, we need to go back to the trivial superrotation vacuum.  In this limit, the Weyl covariant derivatives reduce to covariant derivatives on the celestial sphere. At tree level, we recover the celestial diamond story as presented in~\cite{Pasterski:2021fjn,Pasterski:2021dqe}, illustrated in figure~\ref{fig:gravitondiamond}. The operators appearing  in the expressions of~\cite{Donnay:2021wrk} are plotted at their respective dimensions, also summarized in table~\ref{tab:weights}.  In each case the operators at the bottom corners have dimension $\Delta=3$ while the ones at the top corners have dimension $\Delta=-1$. The construction of the BMS fluxes only requires the bottom half of the memory diamonds. The dashed lines in the figure indicate the descendancy relations that do not appear here.  For comparison with~\cite{Pasterski:2021dqe}, we note that in the linearized case the Schwarzian mode parameterizing the superrotation vacuum can be captured by an operator with the same conformal dimensions as the symmetry parameter $Y$~\cite{Himwich:2019qmj}.

\begin{table}[]
    \centering
    \begin{tabular}{c|c|c|c|c|c}
      ~ &  $P$ & ${\bar J}$ & $\bar {\cal N}^{(0)}$ & $\bar {\cal N}^{(1)}$ & ${\cal C} $ \\
      \hline
        $h$ & $\frac{3}{2}$ & $2$ & $-\frac{1}{2}$ & $-1$ & $-\frac{1}{2} $ \\
        $\bar h$ & $\frac{3}{2}$ & $1$ &  $+\frac{3}{2}$ & $+1$ & $-\frac{1}{2}$ \\
    \end{tabular}
    \caption{Weights of the soft graviton modes appearing in the construction of the loop-corrected celestial stress tensor and their relevant primary descendants.}
    \label{tab:weights}
\end{table}

Both the BMS flux computation and the one-loop exact correction to the subleading soft graviton from amplitudes require that the operator $\bar J_{soft}$ get a contribution that does not descend from the subleading radiative mode $\cal{N}^{(1)}$. The authors of~\cite{He:2017fsb} use the explicit form of the soft factors to match the loop correction to a proposed quadratic operator, effectively considering a double soft limit.  The extrapolate dictionary suggests that we should take the slight mismatch with~\cite{Donnay:2021wrk} seriously, since the transformation laws are sensitive to such modifications.  However, it is worth providing some perspective on the timeline. Issues with non-integrable terms in the definitions of the charges on fixed-$u$ cuts were strategically avoided by sticking to tree level amplitudes in~\cite{Kapec:2014opa}.  The recent work of~\cite{Donnay:2021wrk} builds off lessons learned in the canonical phase space program, while the flux quantities they consider have nice properties precisely because they can be recast in terms of data at the corners where the spacetime has returned to a non-radiative solution.

The Ward identity~\eqref{ward} follows from combining in and out contributions, whereby the corner terms near spatial infinity are canceled by an antipodal matching condition~\cite{Strominger:2017zoo}.  In order for this total charge to vanish, the soft and hard parts need to transform in the same representations. On the one hand, it is straightforward to think of tuning the gravitational coupling to zero to isolate the transformation properties of the hard contribution.  By contrast, the analog for isolating the soft sector is to restrict to non-radiative spacetimes. In the end, the realization of the soft and hard splitting is tied to the leading infrared factorization of the gravitational ${\cal S}$-matrix.

In hindsight, the derivation  of the IR divergent contribution to the loop corrected soft theorem in~\cite{He:2017fsb} arises from the interaction between the soft theorems and the supertranslation dressing. Taking into account the vertex operator construction of~\cite{Himwich:2020rro} indeed remedies the mismatch between~\cite{Donnay:2021wrk} and~\cite{He:2017fsb} and restores a form that transforms as expected under the full asymptotic symmetry group. This provides a new perspective on the question of possible finite corrections to the loop order stress tensor raised in~\cite{He:2017fsb}, since we can phrase this in terms of the ability to construct other boundary graviton modes with the appropriate symmetry transformations. In the end, we see that the mechanism for identifying the soft and hard fluxes
follows the same paradigm as the null state analysis of~\cite{Banerjee:2018fgd,Banerjee:2019aoy,Banerjee:2019tam,Pasterski:2021fjn,Pasterski:2021dqe}, applied to a larger symmetry group. Meanwhile, this example illustrates the power of understanding the underlying bulk perspective when it comes to identifying operators like the loop-corrected stress tensor, whose form is protected by symmetries despite being tricky to pin down unambiguously from the perturbative computation.

\pagebreak

\subsection*{Acknowledgements}
\vspace{-1mm}

Many thanks to Shamik Banerjee, Laurent Freidel, and Herman Verlinde for useful conversations. The author's research has been supported by the Sam B. Treiman Fellowship at the Princeton Center for Theoretical Science.  Research at the Perimeter Institute is supported by the Government of Canada through the Department of Innovation, Science and Industry Canada and by the Province of Ontario through the Ministry of Colleges and Universities.

\appendix

 \bibliographystyle{utphys}
 \bibliography{references}

\end{document}